%% file: Moriond2017_HZZ_Offshell.tex
\pdfoutput=1

\newcommand*{\ATLASLATEXPATH}{latex/}

\documentclass[cernpreprint, texlive=2016, UKenglish]{\ATLASLATEXPATH atlasdoc}

\usepackage[subfigure]{\ATLASLATEXPATH atlaspackage}
\usepackage{\ATLASLATEXPATH atlasbiblatex}

\usepackage{\ATLASLATEXPATH atlascontribute}

\usepackage{\ATLASLATEXPATH atlasphysics}

\graphicspath{{logos/}{figures/}}

\usepackage{Moriond2017_HZZ_Offshell-defs}

\usepackage{multirow}

\input{Moriond2017_HZZ_Offshell-metadata}
\hypersetup{pdftitle={ATLAS document},pdfauthor={The ATLAS Collaboration}}

\begin{document}

\maketitle

\tableofcontents
\clearpage

\section{Introduction}
\label{sec:intro}
\input{./sections/Intro.tex}


\section{ATLAS detector}
\label{sec:detector}
\input{./sections/ATLAS.tex}


\section{Monte Carlo simulation and higher-order theory corrections}
\label{sec:Samples}
\input{./sections/Samples.tex}

\section{\zztollll analysis}
\label{sec:analysis_4l}
\input{./sections/Analysis_4l.tex}

\section{\zztollvv analysis}
\label{sec:analysis_2l2v}
\input{./sections/Analysis_2l2v.tex}
\input{./sections/eventyield_sr.tex}


\section{Systematic uncertainties}
\label{sec:syst}
\input{./sections/Syst.tex}

\section{Results}
\label{sec:result}
\input{./sections/Results.tex}



\FloatBarrier
\section{Conclusion}
\label{sec:conclusion}
\input{./sections/Conclusions.tex}

\section*{Acknowledgements}
\input{acknowledgements/Acknowledgements}



%

\clearpage
\printbibliography

\clearpage
\input{atlas_authlist}

%

\end{document}

%% file: Moriond2017_HZZ_Offshell-metadata.tex
\AtlasTitle{Constraints on off-shell Higgs boson production
and the Higgs boson total width in $\zztollll$ and $\zztollvv$ final states with the ATLAS detector
}

\AtlasAbstract{%
A measurement of off-shell Higgs boson production in the \zztollll\ and \zztollvv\ decay
channels, where $\ell$ stands for either an electron or a muon, is performed using data from
proton--proton collisions at a centre-of-mass energy of $\sqrt{s}=13$~\TeV. The 
data were collected by the ATLAS experiment in 2015 and 2016 at the Large Hadron Collider,
and they correspond to an integrated luminosity of 36.1 fb$^{-1}$.  An observed (expected) upper limit on
the off-shell Higgs signal strength, defined as the event yield normalised to the Standard Model prediction,
of 3.8~(3.4) is obtained at 95\% confidence level~(CL).  Assuming the ratio of the Higgs boson couplings to the
Standard Model predictions is independent of the momentum transfer of the Higgs production mechanism considered in the analysis,
a combination with the on-shell signal-strength measurements yields an observed~(expected) 95\% CL upper limit on the Higgs boson total
width of 14.4~(15.2)~\MeV.  }

\author{The ATLAS Collaboration}

\AtlasRefCode{HIGG-2017-06}

\PreprintIdNumber{CERN-EP-2018-178}

\AtlasJournal{Phys.\ Lett.\ B.}
\AtlasJournalRef{\PLB 786 (2018) 223}
\AtlasDOI{10.1016/j.physletb.2018.09.048}

%% file: sections/Intro.tex
The observation of the Higgs boson by the ATLAS and CMS experiments~\cite{HIGG-2012-27,
CMS-HIG-12-028} at the Large Hadron Collider~(LHC) marks a milestone towards the understanding of the mechanism of
electroweak (EW) symmetry breaking~\cite{Englert:1964et,Higgs:1964pj,Guralnik:1964eu}.  Further
studies of the spin, parity and couplings of the new particle have shown no significant
deviation from the predictions for the Standard Model (SM) Higgs boson~\cite{HIGG-2013-17,
CMS-HIG-14-018, HIGG-2014-06, CMS-HIG-14-009, HIGG-2015-07}.  Efforts to measure the properties of
the Higgs boson are primarily focused on on-shell production.
For a Higgs boson at a mass of 125~\GeV~\cite{HIGG-2014-14, HIGG-2015-07}, the expected natural width of the SM Higgs boson
is $\WHSM\sim4.1$~\MeV~\cite{YR4}.  However, above 125~\GeV\ off-shell production of the Higgs boson has a substantial cross
section at the LHC~\cite{Kauer:2012hd,Caola:2013yja,Campbell:2013una,Campbell:2013wga}, due to the
increased phase space as the vector bosons ($V=W,Z$) and top quark
decay products become on-shell with the increasing energy scale.
This provides an opportunity to study the Higgs boson properties at higher energy scales.
Off-shell production can provide sensitivity to new physics that alters the interactions
between the Higgs boson and other fundamental particles in the high-mass
region~\cite{Englert:2014aca,Cacciapaglia:2014rla,Azatov:2014jga,Ghezzi:2014qpa,Buschmann:2014sia,Gainer:2014hha,Englert:2014ffa,Goncalves:2017iub}.

The measured off-shell event yield from gluon--gluon fusion~(ggF) production normalised to the
SM prediction, where this ratio is referred to as the signal strength $\muoffshell$, can be expressed as
\begin{equation}
\muoffshell = \frac{\xsOFFgg}{\xsOFFggSM} = \kgoff^2 \cdot
\kZoff^2, \nonumber
\end{equation}
where \xsOFFgg{} is the cross section of the off-shell Higgs boson production via ggF
with subsequent decay into a $ZZ$ pair, and $\kgoff$ and $\kZoff$ 
are the off-shell coupling modifiers relative to the SM predictions
associated with the \ggHoff\ production and the \offHtoZZ\ decay, respectively.
The off-shell Higgs boson signal cannot be treated independently of the $gg \to ZZ$ background, 
as sizeable negative interference effects appear~\cite{Kauer:2012hd}. 
The interference term is assumed to be proportional to $\sqrt{\muoffshell}=\kgoff \cdot \kZoff$.
Similarly, $\mu_{\text{on-shell}}$ for the on-shell Higgs boson production via ggF is given by:
\begin{equation}
\muonshell = \frac{\xsONgg}{\xsONggSM} = \frac{\kgon^2\cdot
\kZon^2}{\Gamma_{H}/\Gamma_{H}^{\text{SM}}}, \nonumber
\end{equation}
which depends on the Higgs boson total width $\Gamma_{H}$.
A measurement of the relative off-shell and on-shell event yields,
$\muoffshell/\muonshell$, provides direct information about \WH, if one assumes identical on-shell and
off-shell Higgs boson coupling modifiers~\cite{Campbell:2013una, Passarino:2013bha}.  The above formalism
describing the ratio of off-shell to on-shell cross sections also applies to the vector-boson
fusion~(VBF) production mode.
As in the previous measurement~\cite{HIGG-2014-10}, for a measurement of \WH\ it is necessary to assume that the on-shell
and off-shell coupling modifiers are the same, and for an upper limit that the on-shell coupling modifiers are
not larger than the off-shell couplings.
It is also assumed that any new physics which modifies the off-shell signal strength and the off-shell couplings
does not modify the relative phase of the interfering signal and background processes.
Further, it is assumed that there are neither sizeable kinematic modifications to the off-shell signal nor new
sizeable signals in the search region of this analysis unrelated to an enhanced off-shell signal strength.

The ATLAS and CMS experiments have presented studies of the off-shell production of the Higgs boson using
Run-1 proton--proton ($pp$) collisions data~\cite{HIGG-2014-10,CMS-HIG-14-002,CMS-HIG-14-036,CMS-HIG-14-032}.
ATLAS obtained an observed~(expected) upper limit on the off-shell Higgs boson signal strength~(\muoffshell) in the
range of 5.1--8.6 (6.7--11.0)~\cite{HIGG-2014-10}, using the $ZZ$ and $WW$ channels.
This range is determined by the assumption that the $gg\to ZZ$ and $gg\to WW$ background \kfacs, corresponding to the
ratio of the next-to-leading-order~(NLO) QCD predictions to the leading-order~(LO) predictions,
lie between one-half and twice the value of the \ggSZZWW\ signal \kfac.  An observed~(expected)
95\% confidence level~(CL) upper limit of $\WH < 23 (33)$~\MeV\ was obtained, assuming the
\ggZZWW\ background \kfac\ is equal to the \ggSZZWW\ signal \kfac.
CMS presented a similar study in the $ZZ$ and $WW$ channels, with observed~(expected)
95\% CL upper limit of $\WH < 13 (26)$~\MeV~\cite{CMS-HIG-14-032}.
By comparison, the precision of \WH\ from direct on-shell Higgs boson mass
measurements alone is approximately 1~\GeV~\cite{HIGG-2013-12,CMS-HIG-14-009,CMS-HIG-16-041}, limited
by measurement resolution.

This Letter presents an analysis of off-shell Higgs boson production in the \zztollll\
and \zztollvv\ final states ($\ell=e,\mu$), using \lumi of data collected by the ATLAS detector in
$pp$ collisions at $\sqrt{s}=13~\TeV$.  The off-shell region is defined by requiring the invariant mass of
the $ZZ$ system~($m_{ZZ}$) to be above the on-shell $ZZ$ production threshold, hence well above the Higgs
boson mass, and the on-shell region is defined by a mass window around the 125~\GeV\ resonance.
This analysis adopts the same methodology used in the Run-1 analysis reported in
Ref.~\cite{HIGG-2014-10}.  The analysis for the \zztollll\ final state closely follows the Higgs
boson measurements and high-mass search in the same final state described in
Refs.~\cite{HIGG-2016-25, HIGG-2016-19}.  The off-shell Higgs signal strength is
extracted using a matrix-element discriminant, defined in Section~\ref{sec:analysis_4l},
in a mass region 220~\GeV~$< m_{4\ell} <$~2000~\GeV.  The on-shell signal strength was measured
in the 118~\GeV~$< m_{4\ell}<$~129~\GeV~region in Ref.~\cite{HIGG-2016-25}.  The analysis of
the \zztollvv\ channel, described in Section~\ref{sec:analysis_2l2v},
follows a strategy similar to that used in the search for heavy $ZZ$
resonances described in Ref.~\cite{HIGG-2016-19}.  For this channel, the signal strength is
extracted from the transverse mass distribution in the 250 to 2000~\GeV~range. For off-shell
production of the Higgs boson, the dominant processes of ggF and VBF are
considered. Next-to-next-to-leading-order~(NNLO) QCD and NLO EW corrections are known for the
off-shell signal process \ggS~\cite{Passarino:2013bha}.  More recently, NLO QCD corrections have
also become available for the \ggB\ background and for the signal--background
interference~\cite{Caola2015,Caola2016}, for which additional details are given in
Section~\ref{sec:Samples}.  Given that the QCD corrections for the off-shell signal processes have
only been calculated inclusively in the jet multiplicity, the analysis is performed inclusively in
jet observables and the event selection is designed to minimise the dependence on the momentum of
the $ZZ$ system, which is sensitive to the jet multiplicity.

%% file: sections/ATLAS.tex
The ATLAS experiment is described in Ref.~\cite{PERF-2007-01}. ATLAS is a multipurpose detector with a forward–backward symmetric cylindrical geometry and a solid-angle\footnote{The ATLAS experiment uses a right-handed coordinate system with its origin at the nominal interaction point (IP) in the centre of the detector and the $z$-axis along the beam pipe. The $x$-axis points from the IP to the centre of the LHC ring, and the $y$-axis points upward. Cylindrical coordinates $(r,\phi)$ are used in the transverse plane, $\phi$ being the azimuthal angle around the $z$-axis. The pseudorapidity is defined in terms of the polar angle $\theta$ as $\eta=-\ln\tan(\theta/2)$.} coverage of nearly 4$\pi$.
The inner tracking detector, covering the region $|\eta| <$ 2.5, consists of a silicon pixel detector, a silicon microstrip detector and a straw-tube transition-radiation tracker. The innermost layer of the pixel detector, the insertable B-layer~\cite{ATLAS-TDR-19}, was installed between \RunOne and \RunTwo of the LHC. The inner detector is surrounded by a thin superconducting solenoid providing a $2\,\text{T}$ magnetic field, and by a finely segmented lead/liquid-argon (LAr) electromagnetic calorimeter covering the region $|\eta| <$ 3.2. A steel/scintillator-tile hadronic calorimeter provides coverage in the central region $|\eta| <$ 1.7. The endcap and forward regions, covering the pseudorapidity range 1.5 $< |\eta| <$ 4.9, are instrumented with electromagnetic and hadronic LAr calorimeters, with copper or tungsten as the absorber material.
A muon spectrometer system incorporating large superconducting toroidal air-core magnets surrounds the calorimeters. Three layers of precision wire chambers provide muon tracking in the range $|\eta| <$ 2.7, while dedicated fast chambers are used for triggering in the region $|\eta| <$ 2.4. 
The trigger system is composed of two stages~\cite{TRIG-2016-01}. 
The first stage, implemented with custom hardware, uses information from calorimeters and muon chambers to reduce the event rate from about 40~MHz to a maximum of 100~kHz. The second stage, called the high-level trigger, reduces the data acquisition rate to about 1~kHz on average. The high-level trigger is software-based and runs reconstruction algorithms similar to those used in the offline reconstruction.

%% file: sections/Samples.tex
Monte Carlo~(MC) samples of \ggSBI\ events, which include the SM Higgs boson signal, \ggS{},
the continuum background, \ggB{}, and the signal--background interference
contribution, were generated with the MC generator 
\progname{Sherpa}-v2.2.2~+~\progname{OpenLoops}~\cite{Cascioli:2013gfa,Gleisberg:2008ta,Cascioli:2011va,Denner:2014gla}. 
Matrix elements were calculated for zero jets and one jet at LO
and merged with the \SHERPA parton shower~\cite{Schumann:2007mg}.
The \textsc{NNPDF30NNLO}~\cite{Ball:2014uwa} PDF set was used,
and the QCD renormalisation and factorisation scales were set to $m_{ZZ}$/2.

The \kfac\ for the \ggS{} process is known up to NNLO in QCD as a function of $m_{ZZ}$~\cite{Passarino:2013bha, YR4}.
More recently, a NLO QCD calculation which includes the \ggB\ continuum process has become
available~\cite{Caola2015,Caola2016} allowing $m_{ZZ}$ differential \kfacs\ to be
calculated with an expansion in the inverse top mass~(1/$m_{t}$) below 2$m_{t}$, and assuming a massless-quark approximation
above this threshold. This NLO QCD calculation was used to correct all three components with separate
\kfacs\ computed for the signal \ggS{}~(\KHmZZ), the background \ggB{}~(\KBmZZ) and the
interference~(\KintmZZ).
Since the NNLO QCD correction is only
known differentially in $m_{ZZ}$ for the \ggS{} process and not for all three components in the off-shell region, 
an overall correction is applied by scaling the differential NLO QCD reweighted cross section by an additional factor of 1.2, which is assumed to be the same for the signal, background and interference.
This additional constant
scale factor is justified by 
the constant NNLO to NLO ratio of the QCD predictions
over the data region considered in the analysis.  Using these scaled NLO \kfacs, the cross section for the \ggSBI\ process
with any off-shell Higgs boson signal strength \muoffshell\ can be obtained from a parameterisation
of three SM MC samples: the \ggS\ signal (\sS), the \ggB\ continuum background (\sB) and the full
process with signal, background and interference \ggSBI\ ($\sSBI^\text{SM}$), where the last sample
is required to derive the interference sample:
\begin{eqnarray}
\sSBI(\muoffshell)& =& \muoffshell \cdot 1.2 \cdot \KHmZZ \cdot \sS \nonumber \\
& +&  \sqrt{\muoffshell} \cdot 1.2 \cdot \KintmZZ \cdot \sI  \label{eqn:ggHZZscaling_on_S_B_I_HO} \\
& +&  1.2 \cdot \KBmZZ \cdot \sB, \nonumber \\
\sI &=& \sSBI^\text{SM} -\sS -\sB. \label{eqn:ggHZZscaling_on_I}
\end{eqnarray}

The electroweak $pp\to VV + 2j$ processes containing both the VBF-like events and
events from associated Higgs production with vector bosons~($VH$),
which includes on-shell Higgs boson production, were simulated
using \MGMCatNLO~\cite{Alwall:2014hca} with matrix elements calculated at LO.
The QCD renormalisation and factorisation scales were set
to $m_W$ following the recommendation in Ref.~\cite{LHCHiggsCrossSectionWorkingGroup:2012vm} and
the \textsc{NNPDF23LO} PDF set~\cite{Ball:2012cx} was used.
\progname{Pythia}~8.186~\cite{Sjostrand:2007gs} was used for parton showering and hadronisation,
with the A14 set of tuned parameters for the underlying event~\cite{ATL-PHYS-PUB-2014-021}.
Due to the different $\Gamma_H$ dependence, the on-shell and off-shell Higgs boson production processes are
separated when weighting MC events as in Eqs.~(\ref{eqn:ggHZZscaling_on_S_B_I_HO})
by requiring that the generated Higgs boson mass satisfy $|m_H^\text{gen.}-125~\GeV|<1~\GeV$.
This requirement is fully efficient in selecting the on-shell $VH$ process.
The cross section $\sigma_{pp\rightarrow VV + 2j}(\mu_\text{off-shell})$ for the electroweak $pp\to VV + 2j$ process
for any off-shell Higgs boson signal strength $\mu_\text{off-shell}$ is parameterised in
the same way as for the \ggSBI\ process.

The \qqZZ\ background was simulated with \progname{Sherpa}~v2.2.2, using
the \textsc{NNPDF30NNLO} PDF set for the hard-scattering process. NLO QCD accuracy is achieved in the
matrix-element calculation for 0- and 1-jet final states and LO accuracy for 2- and 3-jet
final states. The merging with the \progname{Sherpa} parton shower was performed using the
\progname{MePs}@NLO prescription.
NLO EW corrections are applied as a function of the particle-level $m_{ZZ}$~\cite{Biedermann:2016yvs,Biedermann:2016lvg}.

The $WW$ and $WZ$ backgrounds were simulated at NLO in QCD using the \POWHEGBOX v2
event generator~\cite{Nason:2013ydw}
with the \progname{CT10NLO} PDF set~\cite{Lai:2010vv} and \progname{Pythia}~8.186 for
parton showering and hadronisation. The non-perturbative effects were modelled with the AZNLO set of tuned parameters~\cite{STDM-2012-23}.
The interference between the \qqZZ\ and $q\bar{q}\to WW$ processes for the $2\ell2\nu$ final state
is found to be negligible and thus is not considered.

Events containing a single $Z$ boson with associated jets~(\Zjets) were simulated using the \SHERPAV{v2.2.1} event generator.
Matrix elements were calculated for up to two partons at NLO and four partons at LO using
the \progname{Comix}~\cite{Gleisberg:2008fv} and \progname{OpenLoops}~\cite{Cascioli:2011va} matrix-element
generators and merged with the \SHERPA parton shower~\cite{Schumann:2007mg} using
the \progname{MePs}@NLO prescription.
The \textsc{NNPDF30NNLO} PDF set was used in conjunction
with dedicated parton-shower tuning developed by the \SHERPA authors. The \Zjets
events are normalised using the NNLO cross sections~\cite{GAVIN20112388}.

The triboson backgrounds $ZZZ$, $WZZ$, and $WWZ$ with fully leptonic decays and at least four prompt charged leptons were modelled
using \SHERPAV{v2.1.1}. The contribution from triboson backgrounds with one $W$ or $Z$ boson decaying hadronically is not included
in the simulation, but the impact on the analysis is found to be negligible.
For the fully leptonic $t\bar{t}+Z$ background, with four prompt charged
leptons originating from the decays of the top quarks and $Z$ boson, \MGMCatNLO was used.
The $t\bar{t}$ background, as well as the single-top and  $Wt$ production, were modelled using \POWHEGBOX v2
interfaced to \PYTHIAV{6.428}~\cite{pythia} with the Perugia 2012~\cite{Skands:2010ak} set of tuned parameters
for parton showering, hadronisation and the underlying event, and to \progname{EvtGen}~v1.2.0~\cite{Lange:2001uf}
for properties of the bottom and charm hadron decays.

The particle-level events produced by each MC event generator were
processed through the ATLAS detector simulation within the \GEANT 4 framework~\cite{GEANT4, SOFT-2010-01}
or the fast detector simulation package Atlfast-II~\cite{SOFT-2010-01}.
Additional $pp$ interactions in the same and nearby bunch crossings (\pu) are included in the simulation.
The \pu\ events were generated using~\progname{Pythia~8} with the A2 set of tuned parameters~\cite{ATL-PHYS-PUB-2012-003}
and the  MSTW2008LO PDF set~\cite{Martin:2009iq}. The simulation samples were weighted to
reproduce the observed distribution of the mean number of interactions per bunch crossing in the data.

%% file: sections/Analysis_4l.tex
The analysis for the \zztollll\ final state closely follows the on-shell Higgs boson measurements
and high-mass search in the same final state described in Refs.~\cite{HIGG-2016-25, HIGG-2016-19},
with the same event reconstruction, trigger and event selections, and background estimation methods.
A matrix-element-based~(ME-based) discriminant computed at LO is constructed to enhance the separation between
the \ggS\ signal and the \ggB\ and \qqZZ\ backgrounds, and this discriminant is subsequently used in a binned
maximum-likelihood fit for the final result.  To minimise the dependence of the \ggZZ\ kinematics on
higher-order QCD effects, the analysis is performed inclusively, ignoring the number of jets in the
events.

The analysis is split into three channels ($4\mu$, $2e2\mu$, $4e$).
Each electron~(muon) must have transverse momentum $\pt > 7~(5)$~\GeV\ and be measured in the
pseudorapidity range $|\eta| < 2.47$ ($|\eta| < 2.7$).  The highest-$\pt$ lepton in the quadruplet
must satisfy $\pt > 20$~\GeV, and the second (third) lepton in $\pt$ order is required to have $\pt >
15$~\GeV\ ($\pt > 10$~\GeV).  Lepton pairs are formed from same-flavour opposite-charge leptons.
For each channel, the quadruplet with a lepton pair whose mass is closest to the $Z$ boson mass is kept.
This pair is referred to as the leading dilepton pair and its invariant mass, $m_{12}$,
is required to be between 50~\GeV\ and 106~\GeV.
The second (subleading) pair is chosen from the remaining leptons
as the pair closest in mass to the $Z$ boson and in the range 50~\GeV~$ <
m_{34} < 115$~\GeV.  The off-shell region is defined as the range 220~\GeV~$< m_{4\ell}
<$~2000~\GeV, while the on-shell region is defined as 118~\GeV~$< m_{4\ell}<$~129~\GeV.

The dominant background in the \zztollll\ channel arises from \qqZZ\ events.  This is modelled using
MC simulation, accurate to NLO QCD and NLO EW corrections as explained in Section~\ref{sec:Samples}.  Other
backgrounds, such as triboson production, $ttV$, \Zjets, and top quark production, constitute less than 2\% of
the total background in the off-shell signal region, and are either taken from simulation or from
dedicated data control regions.

\begin{figure}[!htbp]
\begin{center}
\subfigure[]{\includegraphics[width=0.55\figwidth]{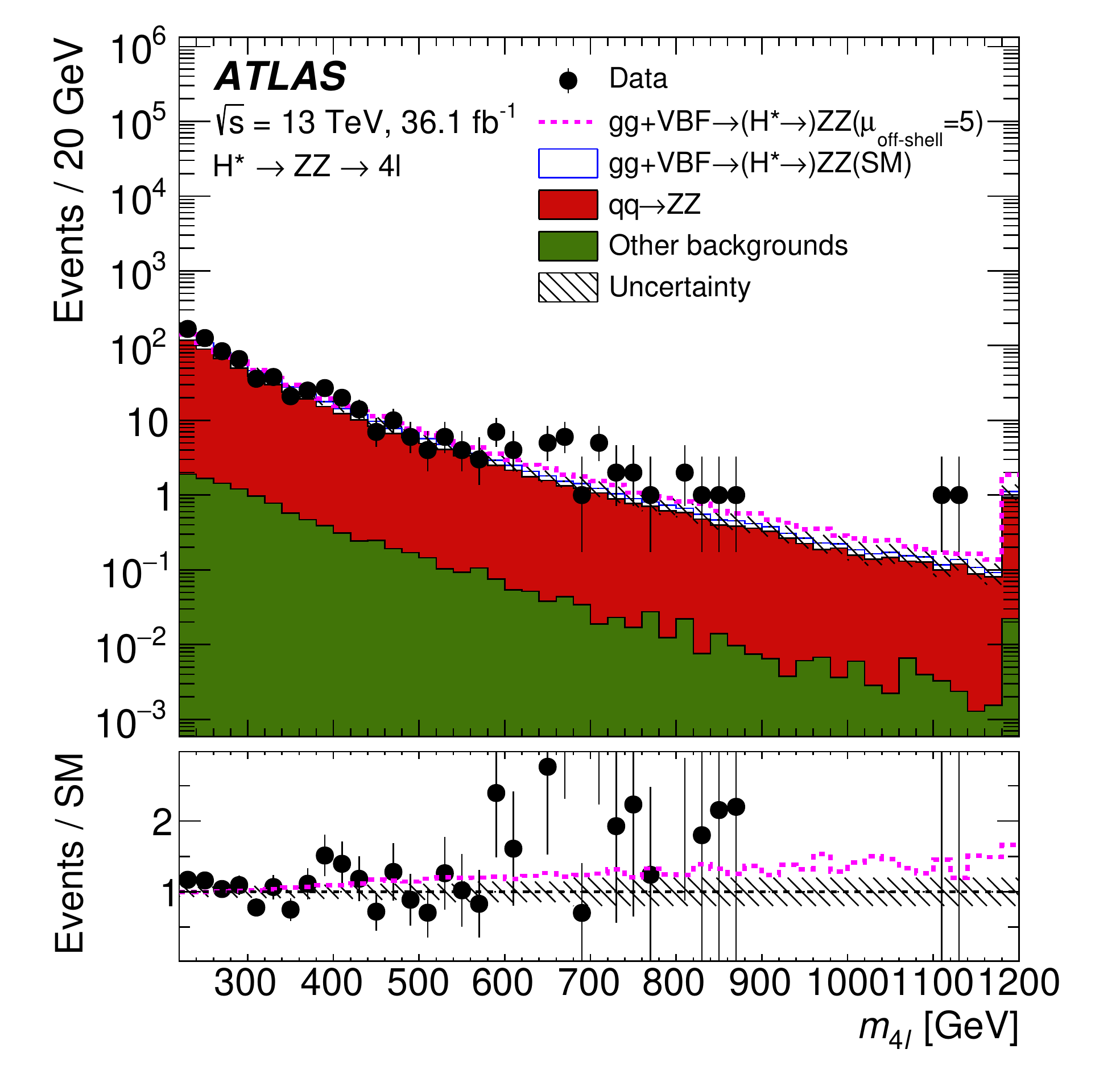}{\label{fig:m4l_data_and_exp:m4l}}}
\subfigure[]{\includegraphics[width=0.55\figwidth]{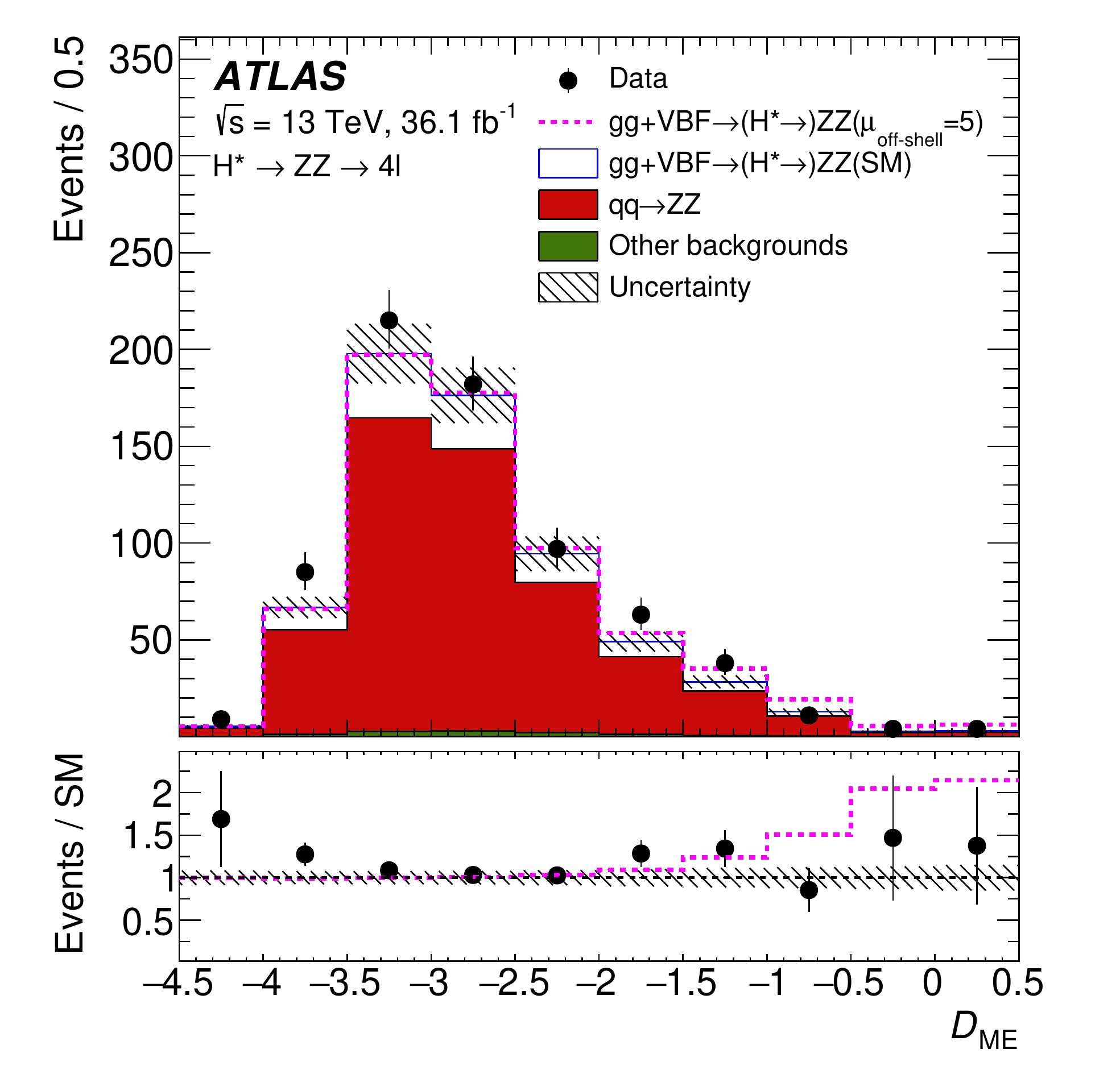}\label{fig:m4l_data_and_exp:ME}}
\caption{Observed distributions in the range 220~\GeV~$< m_{4\ell} <$~2000~\GeV\
for (a) the four-lepton invariant mass $m_{4\ell}$ and (b) the ME-based discriminant \DME\
combining all lepton final states, compared to the expected contributions from the SM including the Higgs boson (stacked).
Events with $m_{4\ell} > 1200$~\GeV\ are included in the last bin of the $m_{4\ell}$ distribution.
The hatched area shows the combined statistical and systematic uncertainties.
The dashed line corresponds to the total expected event yield, including
all backgrounds and the Higgs boson with $\muoffshell=5$.
The ratio plot shows the observed data yield divided by the SM prediction~(black points) as well as the
total expected event yield with $\muoffshell=5$ divided by the SM prediction~(dashed line) in each bin.
}
\label{fig:m4l_data_and_exp}
\end{center}
\end{figure}

Figure~\ref{fig:m4l_data_and_exp:m4l} shows the observed and expected
distributions of $m_{4\ell}$ combining all lepton channels in the off-shell region.
The data are in agreement with the SM predictions,
with two small excesses at $m_{4\ell}$ around 240~\GeV~and 700~\GeV,
each having a significance of about two standard deviations~($2\sigma$),
as evaluated by the high-mass resonance search reported in Ref.~\cite{HIGG-2016-19}.
Table~\ref{tab:yield_cut_data} shows the expected and observed numbers of events in
the signal region and additionally in the 400~\GeV~$ <m_{4\ell}<$~2000~\GeV\ mass range, which is enriched in signal.
The latter mass region was chosen for this table since it is optimal for a counting experiment.

The matrix-element kinematic discriminant fully exploits the event kinematics in the centre-of-mass
frame of the $4\ell$ system. It is computed from eight kinematic observables: the three masses 
$m_{4\ell}$, $m_{12}$ and $m_{34}$, and the leading $Z$ boson production angle and
four decay angles defined in Ref.~\cite{HIGG-2013-01}.
These observables are used to
calculate the matrix elements for the different processes with the \progname{MCFM}
program~\cite{Campbell:2013una} at LO.  The following matrix elements are calculated for each event in the
mass range 220~\GeV~$< m_{4\ell} <$~2000~\GeV:
\begin{itemize}
\item $P_{q\bar{q}}$: the matrix element squared for the $\qqZZ \to 4\ell$ process,
\item $P_{gg}$: the matrix element squared for the $\ggSBI \to 4\ell$ process, which includes the Higgs boson with SM couplings, the continuum background and their interference,
\item $P_H$: the matrix element squared for the $\ggS \to 4\ell$ process without continuum background or interference.
\end{itemize}
The ME-based discriminant is defined as in Ref.~\cite{Campbell:2013una}:
\begin{linenomath}
\begin{equation}
\label{eq:mekd}
\DME = \log_{10}\left({\frac{P_H}{P_{gg}+c\cdot P_{q\bar{q}}}} \right),  \nonumber
\end{equation}
\end{linenomath}
where $c=0.1$ is a constant whose value is chosen to balance the overall cross sections of the
\qqZZ\ and \ggSBI\ processes.
The value of $c$ has a small effect on the analysis sensitivity.
Figure~\ref{fig:m4l_data_and_exp:ME} shows the observed and expected distributions of \DME.
Events with a \DME\ value between $-$4.5 and 0.5 are used for the final result.

%% file: sections/Analysis_2l2v.tex
The analysis in the \zztollvv final state closely follows the one performed to search for \zz resonances~\cite{HIGG-2016-19}.
The reconstruction, identification and selection of electrons, muons, jets, 
$b$-jets and missing transverse momentum are identical while the event selection is optimised for the current analysis.

To discriminate the signal from the background and enhance the sensitivity to off-shell Higgs boson production,
the transverse mass of the \zz system (\mtzz) is used, defined as:
\begin{equation}
\mtzz \equiv \sqrt{ \left[ {\sqrt{\mz^{2} + \left(\pt^{\ell\ell}\right)^2} + \sqrt{\mz^{2} + \left( \met \right)^2}}\: \right]^2 - \left\lvert \vecptll  + \vecmet \right\rvert^2 }, \nonumber
\end{equation}
where $p_{\mathrm{T}}^{\dilep}$ is the transverse momentum of the dilepton system, \mz\ is the mass of the $Z$ boson fixed to
$m_Z=91.187$~\GeV~\cite{ALEPH:2005ab} and $\met$ is the magnitude of the missing transverse momentum $\vecmet$.
The latter is computed as the negative sum of transverse momenta of all the leptons and jets,
as well as the tracks originating from the primary vertex but not associated with any of the leptons or jets, the so-called soft term.

The event selection is designed to minimise the dependence on the \pt of the \zz system, and thus is performed inclusively in number of jets.
First, events with two opposite-charge leptons of the same flavour
are selected with the requirement of $\pt\,> 30$~(20)~\GeV\ for the leading~(sub-leading) lepton.
The dilepton invariant mass $m_{\ell\ell}$ is required to be in the range 76~\GeV~$< m_{\dilep}<106$~\GeV.
Events with additional, loosely identified leptons with $\pt > 7$~\GeV~are rejected to reduce the amount of $WZ$ background.
The two \Zboson bosons originating from the decay of an off-shell Higgs boson are boosted and tend to
be back-to-back in the transverse plane.  A series of selection requirements are applied to reduce
the \Zjets\ background: the two leptons are required to be produced with an angular separation of
$\Delta R\equiv \sqrt{(\Delta \eta)^2 + (\Delta \phi)^2} < 1.8$;
\met is required to be larger than 175~\GeV; the azimuthal angle between
the transverse momentum of the dilepton system and the missing transverse momentum is required to be large,
\begin{math}
\dphizmet > 2.7
\end{math};
$p_{\mathrm{T}}^{\ell\ell}$ is required to be balanced by the \met and jets,
\begin{math}
|p_{\mathrm{T}}^{\mathrm{miss,jet}}-p_{\mathrm{T}}^{\ell\ell}|/p_{\mathrm{T}}^{\ell\ell} < 0.2
\end{math}, where $p_{\mathrm{T}}^{\mathrm{miss, jet}}= |\vec{E}_{\mathrm{T}}^{\mathrm{miss}}+\sum_{\mathrm{jet}}\vec{p}_{\mathrm{T}}^{\mathrm{jet}}|$; and \metht $> 0.33$, where $H_\text{T}$ is the scalar sum of lepton and jet transverse momenta.
Finally, events with a $b$-jet with $\pt\, > \,20$~\GeV\ and $|\eta| \,<$ 2.5, identified by the MV2c10 algorithm~\cite{PERF-2012-04, ATL-PHYS-PUB-2016-012} with 85\% tagging efficiency,
are vetoed to suppress the top quark background.

The dominant backgrounds in the \zztollvv channel consist of \qqZZ\ events, followed by \wz events.
Other background processes with two genuine leptons not directly originating from a $Z$ boson decay include $WW$, $t\bar{t}$, $Wt$ and $Z\rightarrow \tau\tau$.
The remaining background comes from \Zll\ with poorly reconstructed \met, $W$+jets events
with at least one misidentified electron or muon, semileptonic top decays, and multi-jet events.

The \qqZZ\ background is modelled in the same manner as for the \zztollll\ channel.
The \wz{} background is estimated with simulation using a normalisation correction factor extracted
from a dedicated control region~(CR).  This \wz-enriched CR is defined by selecting \Ztoll candidates
with an additional electron or muon with $\pt> 20$~\GeV.  Events with a $b$-jet are rejected to
suppress leptonic $t\bar{t}$ decays and a $m_{\text{T}}(W)\,> 60$~\GeV\ requirement is applied to reduce the \zjets{}
contamination.  The correction factor is then calculated in the CR as the number of data events,
after subtracting the non-$WZ$ contributions, divided by the predicted
$WZ$ yield, and is found to be 1.29. The statistical
uncertainty of the \wz{} estimate is about 2\%, while the systematic uncertainty is estimated to be 5\% from theoretical and
experimental uncertainties in the simulation-based transfer factor between the three-lepton control
region and the two-lepton signal region.

The \nonresll~background, including $WW$, $t\bar{t}$, $Wt$ and $Z\rightarrow \tau\tau$ processes, is estimated from a control sample of $e\mu$ events, in the same manner as in Ref.~\cite{HIGG-2016-19},
except that the $e\mu$ CR is defined by requiring $\met > 120$~\GeV.
The background estimation is performed by extrapolating the result obtained with the relaxed \met requirement to
the SR, extracting the efficiency of the \met selection criteria from MC simulation of the
\nonresll~background.  The \mtzz~distributions for the \nonresll~background are
derived from the data CR and extrapolated to the SR.
The total uncertainty in the \nonresll~estimate is about 40\%, including the statistical uncertainty of the data in the control region, the extrapolation and the method bias estimated from simulation.
The \mtzz~distribution differences between data and simulation are taken as a shape uncertainty~($\sim$10\%).

The \Zll background, expected to be $\sim$2\% of the total background, is estimated from a combination of MC and data-driven
techniques. 
A \Zll enriched CR is defined by reversing the \metht selection. 
Additionally, the $b$-jet veto and the requirement on \dphizmet are removed to allow more data events.
The estimation is performed by extrapolating the number of events observed in the CR,
after subtracting non-$Z$-boson backgrounds, to the SR with a correction factor based on simulation.
The \mtzz distribution for the \Zll\ background is derived from simulation.
The total uncertainty in the \Zll\ estimate is about 50\%~(80\%) for the $ee$~($\mu\mu$) channel,
including the statistical uncertainty of the data in the control region and the extrapolation factor.
The shape difference in \mtzz\ between \Zll\ MC events in the SR and those in the CR is taken into account as a systematic
uncertainty.

Other backgrounds, such as triboson production, $ttV$, \Wjets, and top quark processes other than pair production, constitute only a tiny fraction of the total background in the off-shell signal region, $< 1$\%, and are taken from simulation.
The contribution from the on-shell Higgs production is negligible in the off-shell signal region.

\begin{figure}[htbp]
\begin{center}
\subfigure[]{\includegraphics[width=0.55\figwidth]{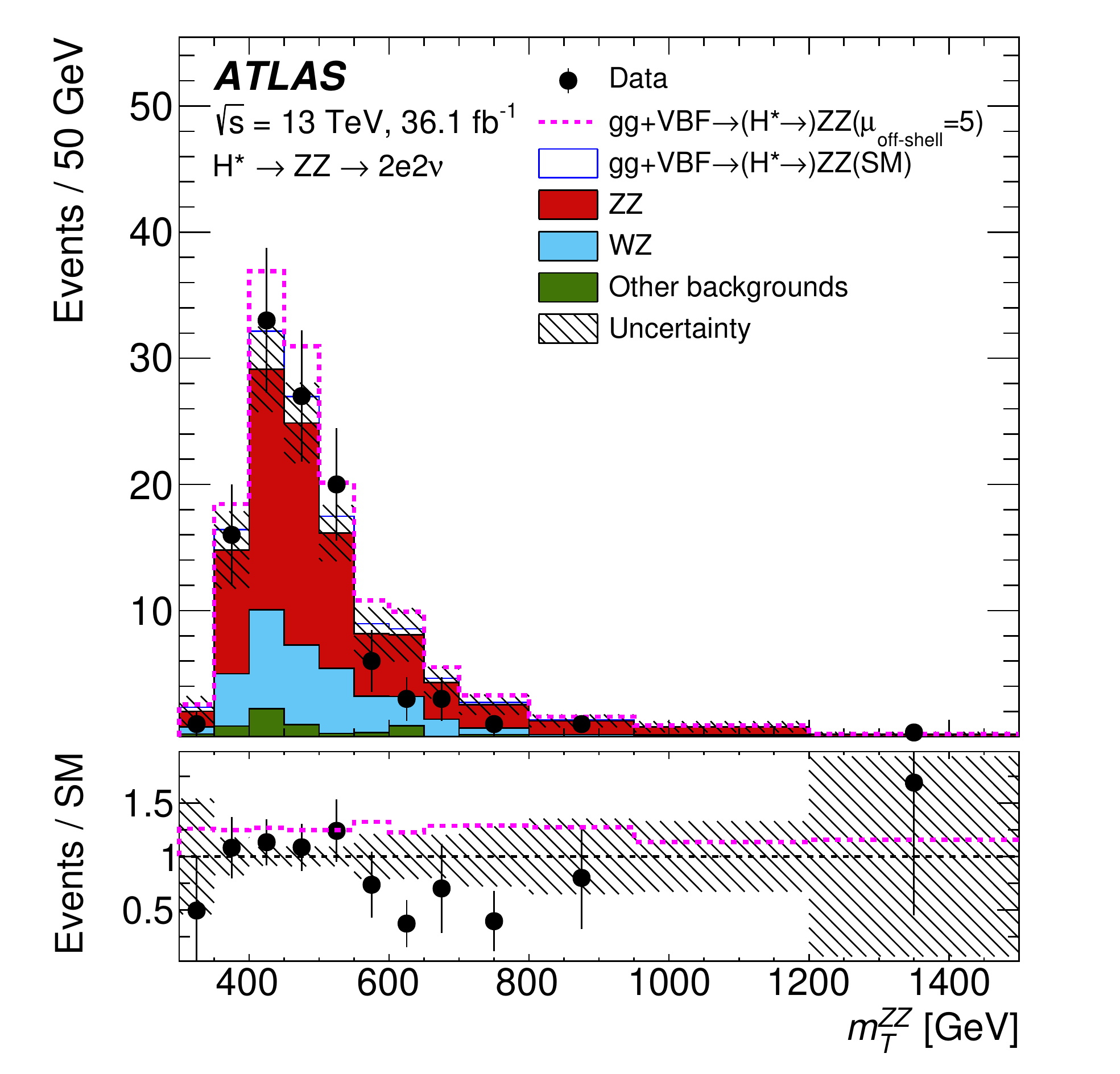}{\label{fig:llvv_mt_ee_data}}}
\subfigure[]{\includegraphics[width=0.55\figwidth]{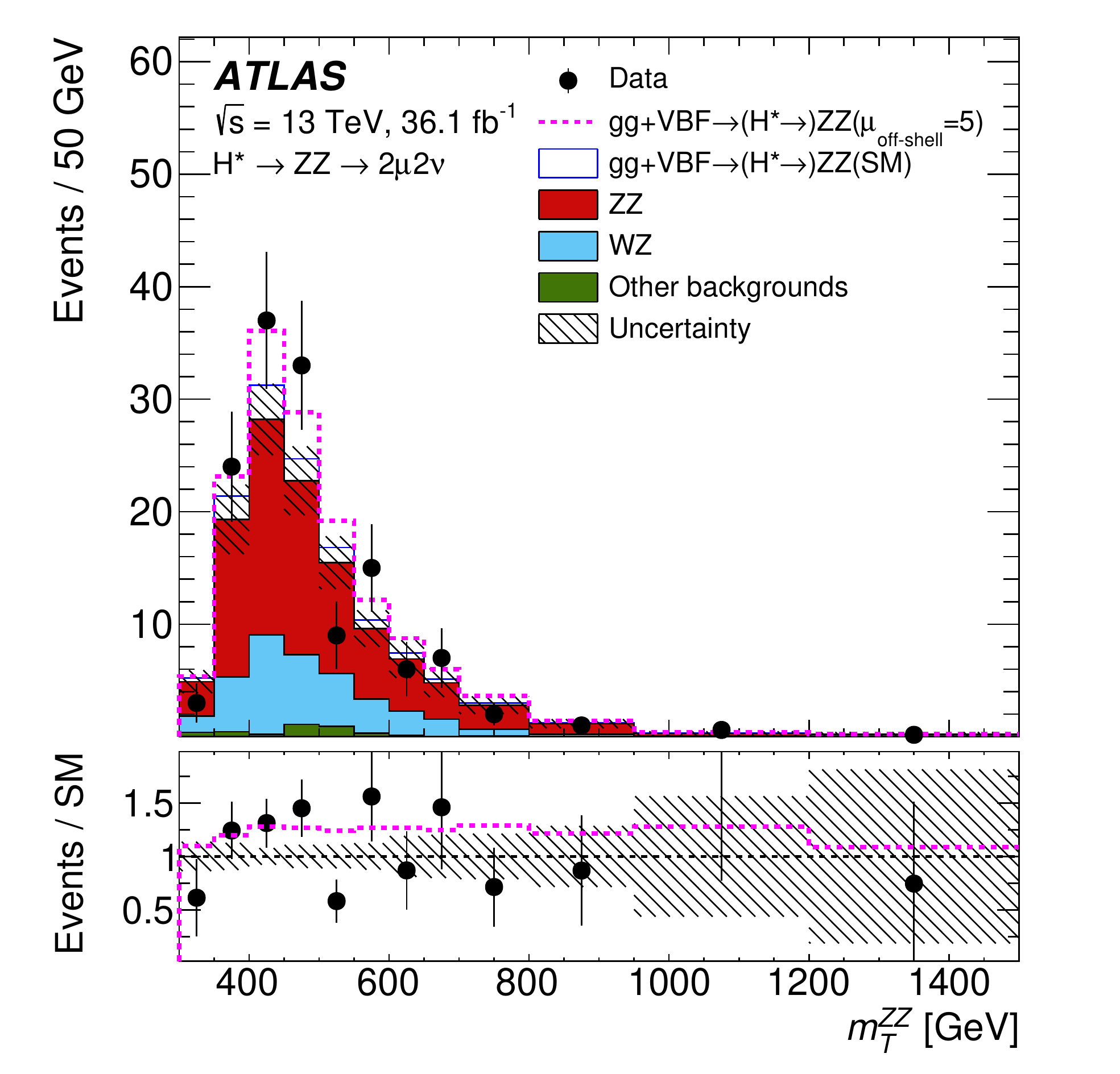}\label{fig:llvv_mt_mm_data}}
\caption{Observed transverse mass \mtzz\ distribution in the (a) $ee$ channel and (b) $\mu\mu$ channel
of the \zztollvv\ off-shell region,
compared to the expected contributions from the SM including the Higgs boson (stacked).
Events with $1500 < \mtzz < 2000$~\GeV\ are included in the last bin of the distribution.
The hatched area shows the combined statistical and systematic uncertainties.
The dashed line corresponds to the total expected event yield, including
all backgrounds and the Higgs boson with $\muoffshell=5$.
The ratio plot shows the observed data yield divided by the SM prediction~(black points) as well as the
total expected event yield with $\muoffshell=5$ divided by the SM prediction~(dashed line) in each bin.
}
\label{fig:SRllvv}
\end{center}
\end{figure}

The expected and observed numbers of events in the signal region for the \zztollvv\ analysis are
summarised in Table~\ref{tab:yield_cut_data}.
Figure~\ref{fig:SRllvv} shows the observed and expected distributions of \mtzz in both the $ee$ and $\mu\mu$ channels in the off-shell region.

%% file: sections/eventyield_sr.tex
\begin{table}[!htb]
\caption{The expected and observed numbers of events in the signal region
for both final states.
For the \zztollll\ analysis, numbers are given for both the signal region and a signal-enriched region which covers the mass range $400~\GeV <m_{4\ell}<2000$~\GeV.
The other backgrounds in the \zztollll\ final state include contributions from \Zjets and top quark processes, while in the \zztollvv\ final state they include contributions from tri-boson production, the \Wjets process, and top quark processes other than pair production.
For the \zztollvv\ analysis, the range 250~\GeV~$< m_{\mathrm{T}}^{ZZ} < 2000$~\GeV\ is considered.
The upper part of the table contains the expected events for the \ggSBI\ and \VBFSBI\ processes which include the Higgs boson signal, background and interference for the SM predictions.
The SM estimates for the signal (S) and background (B) event yields without interference are given in parentheses.
The lower part of the table contains the corresponding predictions for $\mu_\text{off-shell} = 5$.
The uncertainties in the number of expected events include the statistical
uncertainties from MC samples and systematic uncertainties, summed in quadrature.
Empty entries correspond to contributions with event yields smaller than 0.1 events.
}
\label{tab:yield_cut_data}
\centering
\vspace*{0.2cm}
\begin{tabular}{l r@{\,\( \pm \)\,}l r@{\,\( \pm \)\,}l r@{\,\( \pm \)\,}l }

\hline\hline
\multirow{2}{*}{Process}  & \multicolumn{4}{c}{\zztollll} & \multicolumn{2}{c}{\zztollvv} \\
&  \multicolumn{2}{c}{\footnotesize{$m_{4\ell} > 220~\GeV$}} & \multicolumn{2}{c}{\footnotesize{$m_{4\ell} > 400~\GeV$}} 
& \multicolumn{2}{c}{\footnotesize{$m_{\mathrm{T}}^{ZZ} > 250~\GeV$}} \\
\hline
\ggSBI                        & 96   & 15   &  10.6 &  2.0   & 22  & 4    \\

\hspace*{3cm} (\ggS~(S)   & 9.8 & 1.5 &  5.9 &  1.0 & 20.1 & 3.3)  \\
\hspace*{3cm} (\ggB~(B)   & 101 & 16 &  11.8 &  2.2  & 28 & 6)  \\

\VBFSBI                       & 8.29 & 0.34 &  3.07 &  0.13  & 2.83& 0.14 \\
\hspace*{3cm} (\VBFS~(S)  & 1.67 & 0.08 &  1.14 &  0.04  & 5.45 & 0.30)   \\
\hspace*{3cm} (\VBFB~(B)  & 9.9 & 0.4  & 4.17 &  0.18  & 6.92 & 0.35)     \\
$q{\bar q}\to ZZ$              & 520  & 42   &  77   &  8     & 132 & 15   \\
$q\bar{q}\to WZ$              & \multicolumn{2}{c}{-} &  \multicolumn{2}{c}{-}    & 68 & 4     \\
$WW/t\bar{t}/Wt/Z\to\tau\tau$  & \multicolumn{2}{c}{-} &  \multicolumn{2}{c}{-}    & 2.6 & 1.0  \\
\Zjets                         & \multicolumn{2}{c}{-} &  \multicolumn{2}{c}{-}    & 6.0 & 2.8  \\
Other backgrounds              & 14.6 & 0.7 &  2.15 & 0.15 & 1.14 & 0.08 \\
\hline
Total Expected (SM)            & 639  & 60  &  93 &  10  & 234 & 16 \\
Observed  &  \multicolumn{2}{c}{\hspace{-0.3cm}{704}} & \multicolumn{2}{c}{\hspace{-0.3cm}{114}}   &  \multicolumn{2}{c}{\hspace{-0.3cm}{261}} \\
\hline 
Other signal hypothesis & \multicolumn{6}{c}{} \\
\hline
\ggSBIf     &  117 & 18 &  26 &  5  & 61 & 12       \\
\VBFSBIf    & 11.0 & 0.5 & 4.85 &  0.22  & 8.8 & 0.4       \\
\hline\hline

\end{tabular}
\end{table}

%% file: sections/Syst.tex
Systematic uncertainty sources impacting the analysis of both channels can be divided into two categories:
uncertainties in the theoretical description of the signal and background processes
and experimental uncertainties related to the detector or to the reconstruction algorithms.
The largest systematic uncertainties arise from the theoretical uncertainties
in the $gg$-initiated $ZZ$ processes and the \qqZZ\ background process.
The uncertainties from experimental measurements are generally small compared to 
the theoretical uncertainties in this analysis.

The theoretical uncertainties originate from the PDF choice, the missing higher-order corrections,
and the parton-shower modelling. 
The PDF uncertainty corresponds to the 68\% CL variations of the nominal PDF set \textsc{NNPDF30NNLO} for both $q{\bar q} \to ZZ$
and $gg\to (H^\ast\to)ZZ$, as well as the difference from alternative PDF sets.
The alternative PDF sets used are CT10NNLO~\cite{CT10NNLO} and MMHT2014NNLO~\cite{MMHT2014}.
The uncertainty due to PDF is found to be about 3\% in the high-mass region considered.
The uncertainty due to higher-order QCD corrections~(QCD scale uncertainty)
is estimated by varying the renormalisation and factorisation scales independently, 
ranging from a factor of one-half to two.
The uncertainty in the \kfacs\ due to the NLO QCD scale uncertainty is 10--20\% as a function of $m_{ZZ}$ for the 
$gg$-initiated $ZZ$ processes in the probed high-mass region,
and ranges from 5\% to 10\% as a function of $m_{ZZ}$ for the \qqZZ{} background.
The QCD scale uncertainties are treated as correlated among the $gg$-initiated $ZZ$ processes,
and uncorrelated with the $q{\bar q}$-initiated $ZZ$ process.
There are a few additional normalisation uncertainties associated with the NLO \kfacs\
discussed in section\,\ref{sec:Samples}.
In the region below $2m_{t}$, the higher-order corrections are computed with a maximum jet
transverse momentum of $150$~\GeV\ to ensure a good description by the $1/m_{t}$ expansion.  
The default scale uncertainty is therefore doubled for events which have a jet with $\pt >150$~\GeV,
corresponding to about $8\%$ of the events in this region. The scale uncertainty is also increased by 50\% around the $2m_{t}$ threshold, with a Gaussian-smoothed transition decreasing to the default uncertainty
within $50$~\GeV\ of the threshold.
This is intended to allow for possible effects on the \kfac\ which have not been estimated as the
top quark moves on-shell. It is assumed that the 10--20\% NLO QCD scale uncertainty for the
$gg$-initiated $ZZ$ processes covers the assumption of massless loops above the $2m_{t}$
threshold, and as well the uncertainties in the 1.2 scale factor estimated only for the NNLO/NLO signal correction
but also applied to the background and interference components.  
These NLO QCD scale uncertainties are larger than those associated with the NNLO QCD signal uncertainties.
The EW correction uncertainty for \qqZZ\ is evaluated using the same method as in Ref.~\cite{HIGG-2014-10}
and its impact is estimated to be about 1\%.
The parton-shower uncertainty is evaluated by varying parameters in the parton-shower tunes according to
Refs.~\cite{STDM-2012-23, ATL-PHYS-PUB-2014-021} and found to be 2--3\% in normalisation.

The theoretical uncertainties due to the missing higher-order corrections and PDF variations
are small for $VH$-like and VBF-like processes $pp\to ZZ + 2j$; therefore, they are not included in the analysis.

For the \zztollll\ analysis, the same sources of experimental uncertainty as in Ref.~\cite{HIGG-2016-25} are evaluated.
The leading experimental systematic uncertainties are due to the electron and muon
reconstruction and selection efficiency uncertainties, which are smaller than
the uncertainties associated with the theoretical predictions.

Similarly, for the \zztollvv\ channel, the same sources of experimental uncertainty as in Ref.~\cite{HIGG-2016-28}
are evaluated. 
These experimental uncertainties affect the sensitivity of the
\muoffshell\ measurement only at the percent level.

The uncertainty in the combined 2015 and 2016 integrated luminosity is 2.1\%, derived following a methodology similar
to that detailed in Ref.~\cite{DAPR-2013-01}, from a preliminary calibration of the luminosity scale using $x$--$y$
beam-separation scans.
This uncertainty is applied to the normalisation of the signal and also to background contributions whose normalisations
are derived from MC simulations.
A variation in the pile-up reweighting of MC events is included to cover the uncertainty in the ratio of the predicted
and measured inelastic cross sections in Ref.~\cite{STDM-2015-05}.

%% file: sections/Results.tex
The results for the \zztollll\ and \zztollvv\ analyses are first translated into limits on the off-shell
signal strength \muoffshell.
A single off-shell signal-strength parameter is applied for all production modes,
assuming that the ratio of the off-shell production rates via the ggF process to 
those via the VBF process are as predicted in the SM, namely $\muoffshellggF / \muoffshellVBF =1$.
In a second step, the off-shell analyses are combined with the on-shell $ZZ^{*} \to 4 \ell$~\cite{HIGG-2016-22} analysis,
where the on-shell Higgs signal strength is measured to be $\muonshell = 1.28^{+0.21}_{-0.19}$.
The combination with the on-shell analysis is performed with two assumptions that correspond
to different interpretations of the results.
In the first combination, the parameter of interest is the ratio of off-shell to on-shell signal strengths,
which can be interpreted as the Higgs boson width normalised to its SM prediction:
$\muoffshell/\muonshell=\Widthratio$.
This interpretation assumes that the off- and on-shell coupling modifiers are the same for both ggF and
VBF production modes~(i.e.,
$\kgon=\kgoff=\kVon=\kVoff$).
In the second combination, the parameter of interest is the ratio of off-shell to on-shell signal strengths
for the ggF production only, $\Rgg=\muoffshellggF/\muonshellggF$,
which can be interpreted as the ratio of off-shell to on-shell gluon couplings: $\Rgg=\kgoff^2 / \kgon^2$.
In this case the coupling scale factors $\kV=\kVon=\kVoff$
associated with on- and off-shell VBF production and the \HtoZZ\ decay
are assumed to be the same and fitted to the data (profiled).
This also assumes that the total width is equal to the SM prediction.

The statistical analysis is based on the framework described in Refs.~\cite{HIGG-2012-17,Moneta:2010pm,Verkerke:2003ir}.
A binned likelihood function is constructed as a product of Poisson probability terms over all bins of the fit templates
considered. This function depends on the parameter of interest $\mu$, 
corresponding to one of the different interpretations discussed above~(\muoffshell, \Widthratio\ and \Rgg),
and $\theta$, a set of nuisance parameters that encode the effects of systematic
uncertainties on the signal and expected backgrounds, as described in Section~\ref{sec:syst}.
The nuisance parameters are constrained using either Gaussian or log-normal terms.

In the \zztollll\ channel, a binned maximum-likelihood fit to the \DME\ distribution is performed to
extract the limits on $\mu$.
The fit model accounts for signal and background processes, including \ggSBI, \VBFSBI, \qqZZ\ and other backgrounds.
The probability density functions of the signal-related processes
\ggSBI\ and \VBFSBI\ are parameterised as a function of
the off-shell Higgs boson signal strength \muoffshell\
as given in Eqs.~(\ref{eqn:ggHZZscaling_on_S_B_I_HO}) and (\ref{eqn:ggHZZscaling_on_I}).
In the $ZZ\to 2\ell2\nu$ channel, a similar maximum-likelihood fit to the \mtzz\ distribution is performed.
The modelling of the dominant signal and background processes is the same as in the \zztollll\ channel.
The likelihood function for the combination of the \zztollll\ and \zztollvv\ channels is the product of the
Poisson likelihoods of these individual channels.
The main common theoretical and experimental systematic uncertainties are treated as correlated
within different channels.

The PDF uncertainties and uncertainties from higher-order QCD corrections applied to the \qqZZ\ process
are considered correlated between the on-shell and off-shell measurements.
Given the different theoretical computations,
the corresponding uncertainties are considered uncorrelated for the $gg$-initiated $ZZ$ processes between
the on-shell and off-shell measurements, and the impact of such a correlation effect is found to be small.
In addition to the main theoretical uncertainties,
the common experimental systematic uncertainties are treated as correlated between the on-shell and off-shell measurements.

Hypothesis testing and confidence intervals are based on the profile likelihood
ratio~\cite{Cowan:2010js}.  The parameters of interest are different in the various tests,
while the remaining parameters are profiled.
All 95\%~CL upper limits are derived using the \CLs\ method~\cite{CLsmethod},
based on the ratio of one-sided $p$-values:
$R_{\CLs}(\mu) = p_{\mu}/({1-p_{1}})$
where $p_{\mu}$ is the $p$-value for testing a given $\mu=\muoffshell$ or
$\mu=\Widthratio$ (the non-SM hypothesis) and $p_{1}$ is the $p$-value derived
from the same test statistic under the SM hypothesis of $\muoffshell=1$ in the first case and
$\Widthratio =1$ in the second case.%
\footnote{In the context of this analysis the alternative hypothesis is given by the SM value(s)
for all relevant parameters of the fit model.}

\begin{figure}[!ht]
\begin{center}
\subfigure[]{\includegraphics[width=0.55\figwidth]{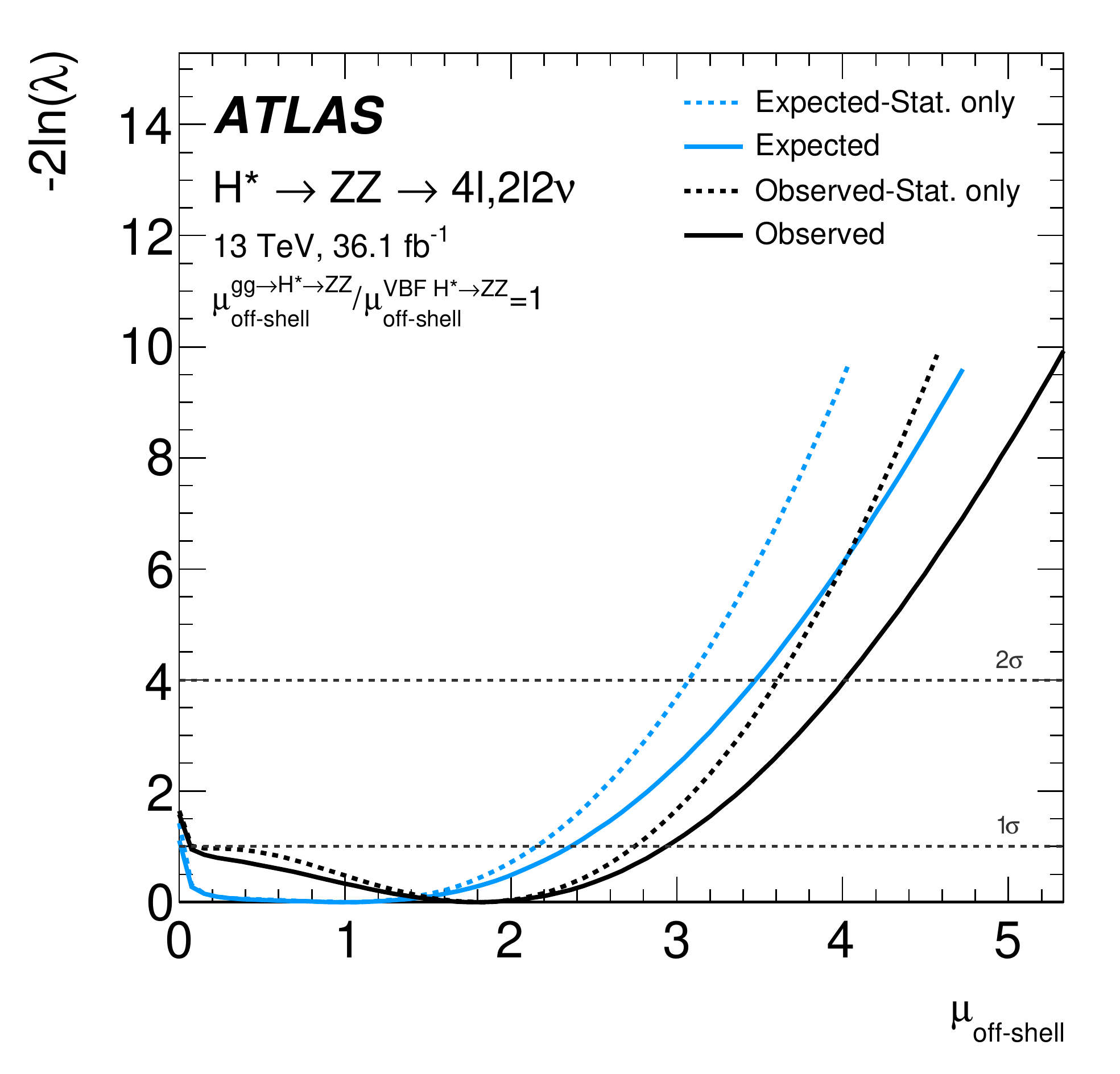}}
\subfigure[]{\includegraphics[width=0.55\figwidth]{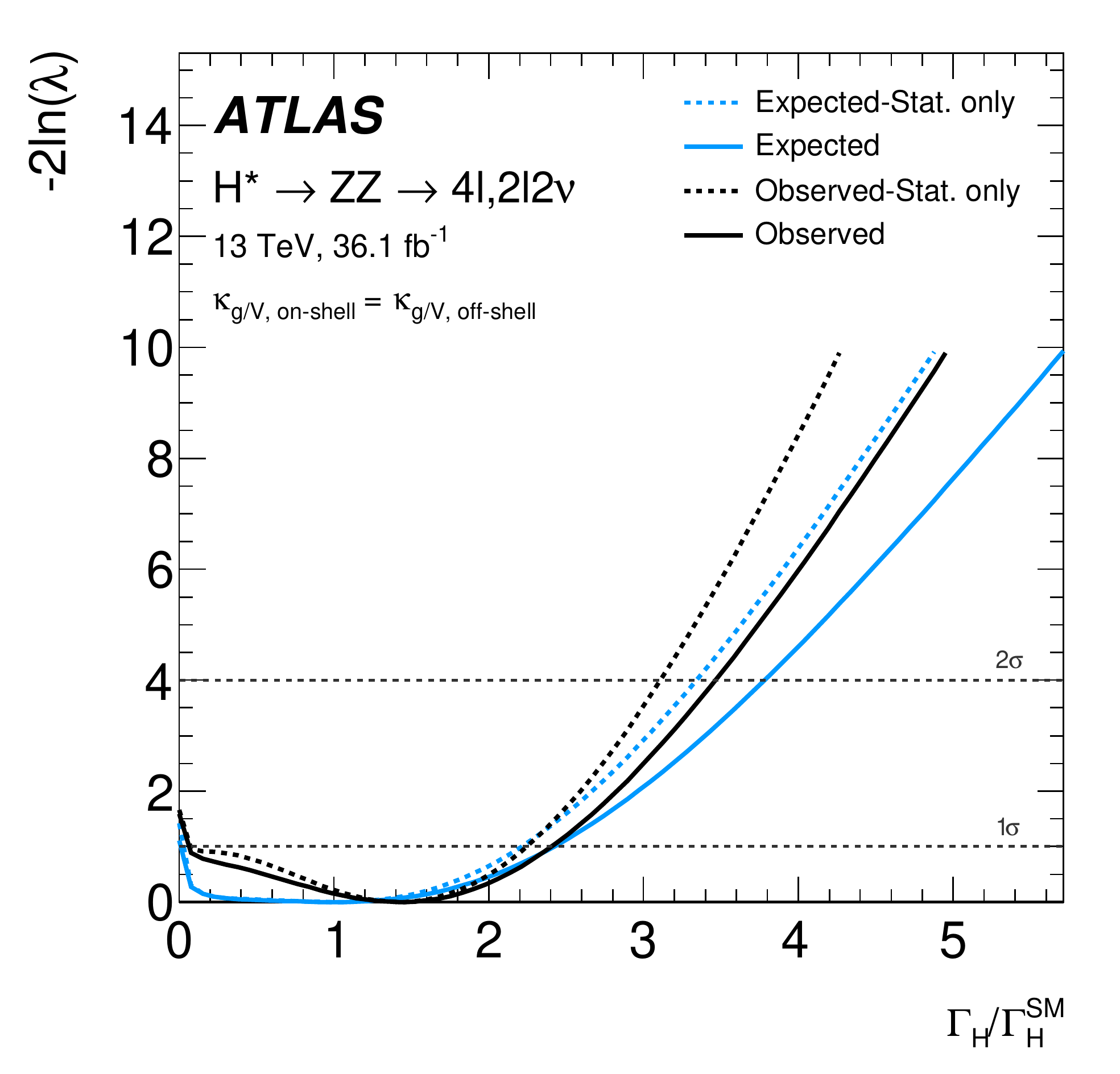}} \\
\subfigure[]{\includegraphics[width=0.55\figwidth]{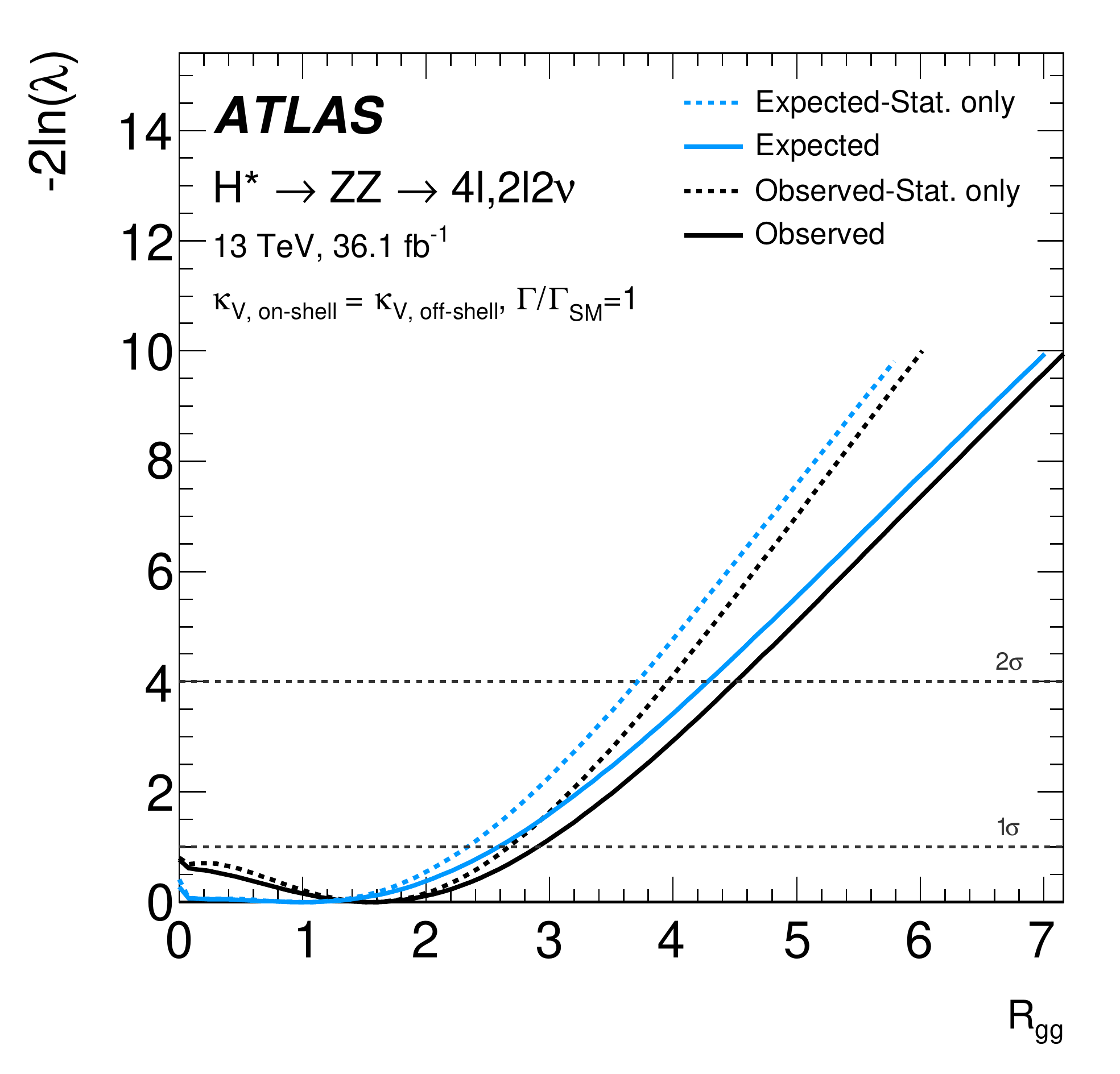}}
\caption{Scan of the negative log-likelihood, $-2\ln\lambda$, for the (a) off-shell Higgs signal strength, \muoffshell\
(b) \Widthratio\ ratio (c) $\Rgg=\kgoff^2 / \kgon^2$.
The solid lower black (upper blue) line represents the observed (expected) value including all systematic uncertainties,
while the dashed lower black (upper blue) line is for the observed (expected) value without systematic uncertainties (lower and upper refer here to the position of the lines in the legend).
The double minimum structure of the scan when the parameter of interest approaches zero is the consequence of the parametrisation
as shown in Eqs.~(\ref{eqn:ggHZZscaling_on_S_B_I_HO}).
}
\label{fig:llh_scan}
\end{center}
\end{figure}

The negative log-likelihood, $-2\ln \lambda$, is scanned as a function of a single parameter of interest,
chosen to be \muoffshell, \Widthratio\ or \Rgg. The results are shown in Figure~\ref{fig:llh_scan} 
for observed and expected values.
The results based on the \CLs\ method for the two individual analyses and their combination are reported
in Table~\ref{tab:limitOFFshell_ZZ}.
As a result of the small data excess observed in the off-shell region, the observed limits on 
\muoffshell\ are less stringent than the expected ones.
The observed (expected) limit on \Widthratio\ is 3.5~(3.7) at the 95\% CL.
Due to the fact that the measured on-shell signal strength \muonshell\ is larger than one~\cite{HIGG-2016-25},
the observed limit on \Widthratio\ is smaller than the expected limit.
The limit on \Widthratio\ can be translated into a limit on the total width of the Higgs boson,
leading to an observed (expected) 95\% CL upper limit on the Higgs boson total width
of 14.4 (15.2)~\MeV.

These results are significantly improved compared to the Run-1 publication~\cite{HIGG-2014-10},
the expected limit being about a factor two better.

If instead of constraining the \qqZZ\ background to the theoretical expectation,
the normalisation is left as a free parameter in the profile likelihood fit, the upper limits on   
\muonshell\ are about 4\% worse in the \zztollll\ channel.
If only the NLO \kfac\ are applied to the SM prediction of the $gg$-initiated $ZZ$ processes,
without the additional NNLO/NLO \kfac\ of 1.2~(Section~\ref{sec:Samples}),
the upper limits on \muoffshell\ and \Widthratio\ are about 10\% worse.

\begin{table}[!htbp]
 \centering
 \caption{The 95\% CL upper limits on \muoffshell, \Widthratio\ and \Rgg.
Both the observed and expected limits are given.
The $1\sigma$~($2\sigma$) uncertainties represent 68\%~(95\%) confidence intervals for the expected limit.
The upper limits are evaluated using the \CLs\ method,
with the SM values as the alternative hypothesis for each interpretation.
}
 \label{tab:limitOFFshell_ZZ}
 \begin{tabular}{llcccc}

\hline\hline
  & & \multirow{2}{*}{Observed} & \multicolumn{3}{c}{Expected} \\
  & & & Median & $\pm1~\sigma$ & $\pm2~\sigma$ \\
\hline
\multirow{3}{*}{\muoffshell} & \zztollll analysis & 4.5  & 4.3 & $[ 3.3, 5.4 ]$ & $[ 2.7, 7.1 ]$ \\
& \zztollvv analysis & 5.3 & 4.4 & $[ 3.4, 5.5 ]$ & $[ 2.8, 7.0 ]$ \\
& Combined           & 3.8 & 3.4 & $[ 2.7, 4.2 ]$ & $[ 2.3, 5.3 ]$ \\
\hline
\Widthratio\  & Combined & 3.5 & 3.7 & $[ 2.9, 4.8 ]$ & $[ 2.4, 6.5 ]$ \\
\hline
\Rgg\         & Combined & 4.3 & 4.1 & $[ 3.3, 5.6 ]$ & $[ 2.7, 8.2 ]$ \\
\hline\hline

 \end{tabular}
\end{table}

The impact of the various systematic uncertainties on the expected limit in the \muoffshell\ fit are
listed in Table~\ref{tab:nuisance_offShell_ZZ_comb}.
The values in this table were derived by fixing all the nuisance parameters associated with the systematic
uncertainties to the values derived from the SM-conditional fit to the data, with the exception of the one
under study. The uncertainties with the largest impact on the sensitivity
of \muoffshell\ are the theoretical uncertainties of the $gg$- and $q{\bar q}$-initiated $ZZ$ processes.

\begin{table}[!ht]
\begin{center}
\caption{The expected 95\% CL upper limit on \muoffshell\
with a ranked listing of the impact of the leading systematic uncertainty individually, comparing with
no systematic uncertainty or all systematic uncertainties.
The upper limits are evaluated using the \CLs\ method.}
\label{tab:nuisance_offShell_ZZ_comb}
\begin{tabular}{lccc}
\hline
\hline
\multirow{2}{*}{Systematic uncertainty} &  \multicolumn{3}{c}{95\% CL upper limit on \muoffshell}  \\
 & \zztollll & \zztollvv &  Combined \\
\hline

QCD scale \qqZZ\                            & 4.2        & 3.9       & 3.2 \\
QCD scale \ggSBI\                           & 4.2        & 3.6       & 3.1 \\
Luminosity                                  & 4.1        & 3.5       & 3.1 \\
Remaining systematic uncertainties          & 4.1        & 3.5       & 3.0 \\
\hline                                                         
All~systematic uncertainties                & 4.3        & 4.4       & 3.4 \\
\hline                                                         
No~systematic uncertainties                 & 4.0        & 3.4       & 3.0 \\
\hline
\hline

\end{tabular}
\end{center}
\end{table}

%% file: sections/Conclusions.tex
A determination of the off-shell Higgs boson signal strength in the \zztollll\ and \zztollvv\
final states and their combination is presented.
The result is based on $pp$ collision data collected by the ATLAS experiment at the LHC,
corresponding to an integrated luminosity of \lumi\ at a collision energy of $\sqrt{s}=13~\TeV$.
Using the  \CLs\ method, the observed~(expected) 95\% confidence level (CL) upper limit on the off-shell signal strength is 
3.8~(3.4).
Assuming the ratio of the relevant Higgs boson couplings to the SM predictions are constant
with energy from on-shell production to the high-mass range considered in this analysis,
a combination with the on-shell measurements yields an observed~(expected) 95\% CL upper limit on the
Higgs boson total width of 14.4~(15.2)~\MeV.

Assuming that the total width of the Higgs boson is as expected in the SM, 
and the coupling scale factors associated with on- and off-shell VBF production and the \HtoZZ\ decay are the same,
the same combination can be interpreted as a limit on the ratio of the off-shell
to the on-shell couplings to gluons $\Rgg=\kgoff^2 / \kgon^2$.
An observed (expected) limit of 4.3~(4.1) at 95\% CL on \Rgg\ is obtained.

%% file: acknowledgements/Acknowledgements.tex

We thank CERN for the very successful operation of the LHC, as well as the
support staff from our institutions without whom ATLAS could not be
operated efficiently.

We acknowledge the support of ANPCyT, Argentina; YerPhI, Armenia; ARC, Australia; BMWFW and FWF, Austria; ANAS, Azerbaijan; SSTC, Belarus; CNPq and FAPESP, Brazil; NSERC, NRC and CFI, Canada; CERN; CONICYT, Chile; CAS, MOST and NSFC, China; COLCIENCIAS, Colombia; MSMT CR, MPO CR and VSC CR, Czech Republic; DNRF and DNSRC, Denmark; IN2P3-CNRS, CEA-DRF/IRFU, France; SRNSFG, Georgia; BMBF, HGF, and MPG, Germany; GSRT, Greece; RGC, Hong Kong SAR, China; ISF, I-CORE and Benoziyo Center, Israel; INFN, Italy; MEXT and JSPS, Japan; CNRST, Morocco; NWO, Netherlands; RCN, Norway; MNiSW and NCN, Poland; FCT, Portugal; MNE/IFA, Romania; MES of Russia and NRC KI, Russian Federation; JINR; MESTD, Serbia; MSSR, Slovakia; ARRS and MIZ\v{S}, Slovenia; DST/NRF, South Africa; MINECO, Spain; SRC and Wallenberg Foundation, Sweden; SERI, SNSF and Cantons of Bern and Geneva, Switzerland; MOST, Taiwan; TAEK, Turkey; STFC, United Kingdom; DOE and NSF, United States of America. In addition, individual groups and members have received support from BCKDF, the Canada Council, CANARIE, CRC, Compute Canada, FQRNT, and the Ontario Innovation Trust, Canada; EPLANET, ERC, ERDF, FP7, Horizon 2020 and Marie Sk{\l}odowska-Curie Actions, European Union; Investissements d'Avenir Labex and Idex, ANR, R{\'e}gion Auvergne and Fondation Partager le Savoir, France; DFG and AvH Foundation, Germany; Herakleitos, Thales and Aristeia programmes co-financed by EU-ESF and the Greek NSRF; BSF, GIF and Minerva, Israel; BRF, Norway; CERCA Programme Generalitat de Catalunya, Generalitat Valenciana, Spain; the Royal Society and Leverhulme Trust, United Kingdom.

The crucial computing support from all WLCG partners is acknowledged gratefully, in particular from CERN, the ATLAS Tier-1 facilities at TRIUMF (Canada), NDGF (Denmark, Norway, Sweden), CC-IN2P3 (France), KIT/GridKA (Germany), INFN-CNAF (Italy), NL-T1 (Netherlands), PIC (Spain), ASGC (Taiwan), RAL (UK) and BNL (USA), the Tier-2 facilities worldwide and large non-WLCG resource providers. Major contributors of computing resources are listed in Ref.~\cite{ATL-GEN-PUB-2016-002}.

%% file: atlas_authlist.tex
 
\begin{flushleft}
{\Large The ATLAS Collaboration}

\bigskip

M.~Aaboud$^\textrm{\scriptsize 34d}$,    
G.~Aad$^\textrm{\scriptsize 99}$,    
B.~Abbott$^\textrm{\scriptsize 124}$,    
O.~Abdinov$^\textrm{\scriptsize 13,*}$,    
B.~Abeloos$^\textrm{\scriptsize 128}$,    
D.K.~Abhayasinghe$^\textrm{\scriptsize 91}$,    
S.H.~Abidi$^\textrm{\scriptsize 164}$,    
O.S.~AbouZeid$^\textrm{\scriptsize 39}$,    
N.L.~Abraham$^\textrm{\scriptsize 153}$,    
H.~Abramowicz$^\textrm{\scriptsize 158}$
% ...
\bigskip
\\

% ...
$^{13}$Institute of Physics, Azerbaijan Academy of Sciences, Baku; Azerbaijan.\\
% ...
$^{34}$$^{(a)}$Facult\'e des Sciences Ain Chock, R\'eseau Universitaire de Physique des Hautes Energies - Universit\'e Hassan II, Casablanca;$^{(b)}$Centre National de l'Energie des Sciences Techniques Nucleaires (CNESTEN), Rabat;$^{(c)}$Facult\'e des Sciences Semlalia, Universit\'e Cadi Ayyad, LPHEA-Marrakech;$^{(d)}$Facult\'e des Sciences, Universit\'e Mohamed Premier and LPTPM, Oujda;$^{(e)}$Facult\'e des sciences, Universit\'e Mohammed V, Rabat; Morocco.\\
% ...
$^{39}$Niels Bohr Institute, University of Copenhagen, Copenhagen; Denmark.\\
% ...
$^{91}$Department of Physics, Royal Holloway University of London, Egham; United Kingdom.\\
% ...
$^{99}$CPPM, Aix-Marseille Universit\'e, CNRS/IN2P3, Marseille; France.\\
% ...
$^{124}$Homer L. Dodge Department of Physics and Astronomy, University of Oklahoma, Norman OK; United States of America.\\
% ...
$^{128}$LAL, Universit\'e Paris-Sud, CNRS/IN2P3, Universit\'e Paris-Saclay, Orsay; France.\\
% ...
$^{153}$Department of Physics and Astronomy, University of Sussex, Brighton; United Kingdom.\\
% ...
$^{158}$Raymond and Beverly Sackler School of Physics and Astronomy, Tel Aviv University, Tel Aviv; Israel.\\
% ...
$^{164}$Department of Physics, University of Toronto, Toronto ON; Canada.\\
% ...
$^{a}$ Also at Borough of Manhattan Community College, City University of New York, NY; United States of America.\\
$^{b}$ Also at Centre for High Performance Computing, CSIR Campus, Rosebank, Cape Town; South Africa.\\
$^{c}$ Also at CERN, Geneva; Switzerland.\\
$^{d}$ Also at CPPM, Aix-Marseille Universit\'e, CNRS/IN2P3, Marseille; France.\\
$^{e}$ Also at D\'epartement de Physique Nucl\'eaire et Corpusculaire, Universit\'e de Gen\`eve, Gen\`eve; Switzerland.\\
% ...
$^{*}$ Deceased

\end{flushleft}
